\documentclass[11pt,a4paper]{article}
\usepackage{epsfig}
\usepackage{amssymb}
\usepackage{here}
\usepackage{cite}
\usepackage{rotating}
\begin{document}
%%%\psdraft
%=======================================================================
%       Parameters for the title page
%=======================================================================
%
%  EP number, Date and Version
%
\newcommand{\PPEnum}    {CERN-EP-2002-045}
\newcommand{\PRnum}     {OPAL PR-xxx}
\newcommand{\PNnum}     {OPAL Physics Note PN-xxx}
\newcommand{\TNnum}     {OPAL Technical Note TN-xxx}
\newcommand{\Date}      {1 July 2002}
\newcommand{\Author}    {M.~J.~Oreglia}
\newcommand{\MailAddr}  {Mark.Oreglia@CERN.ch}
\newcommand{\EdBoard}   {E. Duchovni, M. Gruw\'e, D. Karlen, P. M\"attig}
\newcommand{\DraftVer}  {FINAL DISPATCH}
\newcommand{\DraftDate} {\Date}
\newcommand{\TimeLimit} {{\bf Friday, 28 June, 17h00 Geneva time}}

%=======================================================================
%       Parameters affecting the  appearance
%=======================================================================
\def\toprule{\noalign{\hrule \medskip}}
\def\midrule{\noalign{\medskip\hrule }}
\def\botrule{\noalign{\medskip\hrule }}
\setlength{\parskip}{\medskipamount}

%=======================================================================
%       Define symbols
%=======================================================================
%
\newcommand{\general}  {{\it general}}
\newcommand{\SM}  {{Standard Model}}

\newcommand{\ellell}   {\ell^+ \ell^-}
\newcommand{\ee}       {\mbox{${\mathrm{e}}^+ {\mathrm{e}}^-$}}
\newcommand{\epem}     {{\mathrm e}^+ {\mathrm e}^-}
\newcommand {\mm}      {\mu^+ \mu^-}
\newcommand{\mupair}   {\mbox{$\mu^+\mu^-$}}
\newcommand{\nunu}     {\nu \bar{\nu}}
\newcommand{\tautau}   {\mbox{$\tau^+\tau^-$}}
\newcommand{\qqbar}    {{\mathrm q}\bar{\mathrm q}}
\newcommand{\qpair}    {\mbox{${\mathrm q}\overline{\mathrm q}$}}
\newcommand{\ff}       {{\mathrm f} \bar{\mathrm f}}
\newcommand{\gaga}     {\gamma\gamma}
\newcommand{\WW}       {{\mathrm W}^+{\mathrm W}^-}
\newcommand{\ZZ}       {{\mathrm Z}^{0}{\mathrm Z}^{0}}
\newcommand{\Mrec}      {M_{\mrm{recoil}}}
\newcommand{\mgg}       {$m_{\gamma \gamma}$}
\newcommand{\mdip}      {m_{\gamma \gamma}}
\newcommand{\MZ}        {M_{\mathrm Z}}
\newcommand{\MH}        {M_{\mathrm H}}
\newcommand{\MX}        {M_{\mathrm{X}}}
\newcommand{\MY}        {M_{\mathrm{Y}}}
\newcommand {\ho}        {\mbox{$\mathrm{h}^{0}$}}
\newcommand {\Ao}        {\mbox{$\mathrm{A}^{0}$}}
\newcommand {\Ho}        {\mbox{$\mathrm{H}^{0}$}}
\newcommand {\Zo}        {\mbox{$\mathrm{Z}^{0}$}}
\newcommand {\Zboson}        {{\mathrm Z}^{0}}
\newcommand {\Wpm}           {{\mathrm W}^{\pm}}
\newcommand {\hboson}        {{\mathrm h}^{0}}
\newcommand{\ggjj}{\mbox{$\gaga\;{\rm jet-jet}$ }}
\newcommand{\ggqq}{\mbox{$\gaga {\rm q\bar{q}}$ }}
\newcommand{\ggll}{\mbox{$\gaga\ell^{+}\ell^{-}$}}
\newcommand{\ggnn}{\mbox{$\gaga\nu\bar{\nu}$}}
\newcommand{\degree}    {^\circ}
\newcommand{\roots}       {\sqrt{s}}
\newcommand{\Ecm}         {\mbox{$E_{\mathrm{cm}}$}}
\newcommand{\Egam}        {\mbox{$E_{\gamma}$}}
\newcommand{\EgamA}        {\mbox{$E_{\gamma1}$}}
\newcommand{\EgamB}        {\mbox{$E_{\gamma2}$}}
\newcommand{\Ebeam}       {E_{\mathrm{beam}}} 
\newcommand{\ipb}         {\mbox{pb$^{-1}$}}
\newcommand{\Evis}      {\mbox{$E_{\mathrm{vis}}$}}
\newcommand{\Rvis}      {\mbox{$R_{\mathrm{vis}}$}}
\newcommand{\Mvis}      {\mbox{$M_{\mathrm{vis}}$}}
\newcommand{\Rbal}      {\mbox{$R_{\mathrm{bal}}$}}
%-------
%  etc
%-------
\newcommand{\onecol}[2] {\multicolumn{1}{#1}{#2}}
\newcommand{\colcen}[1] {\multicolumn{1}{|c|}{#1}}
\newcommand{\ra}        {\rightarrow}   % \to can be used as well
\newcommand{\ov}        {\overline}   
\def\mrm       {\mathrm}
%
%
% Def. fuer groesser-ungefaehr:
\newcommand{\gsim}{\;\raisebox{-0.9ex}
           {$\textstyle\stackrel{\textstyle >}{\sim}$}\;}
% Def. fuer kleiner-ungefaehr:
%\newcommand{\lsim}{\;\raisebox{-0.9ex}{\textstyle\stackrel{\textstyle$<$}{$\sim$}}\;}
%\def\lsim{\mathrel{\raise.3ex\hbox{$<$\kern-.75em\lower1ex\hbox{$\sim$}}}}
%
%----------------------------
%  Bibliographic references
%----------------------------
%
%     Journal names
%
\newcommand{\PhysLett}  {Phys.~Lett.}
\newcommand{\PRL}       {Phys.~Rev.\ Lett.}
\newcommand{\PhysRep}   {Phys.~Rep.}
\newcommand{\PhysRev}   {Phys.~Rev.}
\newcommand{\NPhys}     {Nucl.~Phys.}
\def\NIM                {\mbox{Nucl. Instr. Meth.}}
\newcommand{\NIMA}[1]   {\NIM\ {\bf A{#1}}}
\newcommand{\IEEENS}    {IEEE Trans.\ Nucl.~Sci.}
% journals
\newcommand{\ZPhysC}[1]    {Z. Phys. {\bf C#1}}
\newcommand{\EurPhysC}[1]    {Eur. Phys. J. {\bf C#1}}
\newcommand{\PhysLettB}[1] {Phys. Lett. {\bf B#1}}
\newcommand{\CPC}[1]      {Comp.\ Phys.\ Comm.\ {\bf #1}}
\def\etal{\mbox{{\it et al.}}}
%
%     Collaboration names
%
\newcommand{\OPALColl}  {OPAL Collab.}
\newcommand{\JADEColl}  {JADE Collab.}
%
%=======================================================================
%       Title Page
%=======================================================================

%-----------------------------------------------------------------------
\begin{titlepage}
\begin{center}{\large   EUROPEAN ORGANIZATION FOR NUCLEAR RESEARCH
}\end{center}\bigskip
\begin{flushright}
       \PPEnum  \\ \DraftDate
\end{flushright}
\bigskip\bigskip\bigskip\bigskip\bigskip
%
%     Main title
%
\begin{center}{\huge\bf\boldmath Search for Associated Production\\ 
                                 of Massive States\\
                                Decaying into Two Photons\\ 
                                in \ee\ Annihilations at $\roots=88-209$~GeV\\ }
\end{center}\bigskip\bigskip
\begin{center}{\LARGE The OPAL Collaboration
}\end{center}\bigskip\bigskip
%
%     Abstract
%
\bigskip\begin{center}{\large  Abstract}\end{center}
A search is performed for production of short-lived particles in
$\epem \ra \mrm {X Y}$, with $\mrm X \ra \gaga$ and $\mrm Y \ra \ff $,
for scalar $\mrm X$ and scalar or vector $\mrm Y$. 
Model-independent limits in the range of 25-60 femtobarns
are presented on 
$\sigma(\epem\ra \mrm {X Y})\times B({\mrm X} \ra \gaga)\times B(\mrm Y \ra \ff)$
for centre-of-mass energies in the range $205-207$~GeV. 
The data from all LEP centre-of-mass energies
$88-209$~GeV are also interpreted in the context of fermiophobic Higgs
boson models, for which a lower mass limit of 105.5~GeV is obtained
for a ``benchmark'' fermiophobic Higgs boson.
\bigskip\bigskip\bigskip\bigskip

\begin{center}{\large
%\DraftVer \\
%Editorial Board: \EdBoard \\
%\bigskip\bigskip
%Please send your comments to \MailAddr \\
%No later than \TimeLimit \\ 
\bigskip\bigskip
(Submitted to Physics Letters B) \\

}\end{center}

\end{titlepage}
%====================================================================================
\begin{center}{\Large        The OPAL Collaboration
}\end{center}\bigskip
\begin{center}{
%begin authorlist PLEASE DO NOT DELETE THIS COMMENT
G.\thinspace Abbiendi$^{  2}$,
C.\thinspace Ainsley$^{  5}$,
P.F.\thinspace {\AA}kesson$^{  3}$,
G.\thinspace Alexander$^{ 22}$,
J.\thinspace Allison$^{ 16}$,
P.\thinspace Amaral$^{  9}$, 
G.\thinspace Anagnostou$^{  1}$,
K.J.\thinspace Anderson$^{  9}$,
S.\thinspace Arcelli$^{  2}$,
S.\thinspace Asai$^{ 23}$
% ...
%end authorlist PLEASE DO NOT DELETE THIS COMMENT
}\end{center}\bigskip
\bigskip
%begin institutes
$^{  1}$School of Physics and Astronomy, University of Birmingham,
Birmingham B15 2TT, UK
\newline
$^{  2}$Dipartimento di Fisica dell' Universit\`a di Bologna and INFN,
I-40126 Bologna, Italy
\newline
$^{  3}$Physikalisches Institut, Universit\"at Bonn,
D-53115 Bonn, Germany
\newline
% ...
$^{  5}$Cavendish Laboratory, Cambridge CB3 0HE, UK
\newline
% ...
$^{  9}$Enrico Fermi Institute and Department of Physics,
University of Chicago, Chicago IL 60637, USA
\newline
% ...
$^{ 16}$Department of Physics, Schuster Laboratory, The University,
Manchester M13 9PL, UK
\newline
% ...
$^{ 22}$Department of Physics and Astronomy, Tel Aviv University,
Tel Aviv 69978, Israel
\newline
$^{ 23}$International Centre for Elementary Particle Physics and
Department of Physics, University of Tokyo, Tokyo 113-0033, and
Kobe University, Kobe 657-8501, Japan
\newline
% ...
$^{  e}$ and Department of Experimental Physics, Lajos Kossuth University,
 Debrecen, Hungary
\newline
$^{  f}$ and MPI M\"unchen
\newline
% ...
$^{  h}$ now at University of Liverpool, Dept of Physics,
Liverpool L69 3BX, UK
\newline
% ...
%end notes
%====================================================================================
\newpage
%=======================================================================
\section{Introduction}
\label{sec:intro}
%=======================================================================
This paper presents the results of two types of search 
for the production of a di-photon system recoiling from 
another massive scalar or vector object.
The searches are sensitive to the processes 
$\epem \ra \mrm {X Y}$, with $\mrm X \ra \gaga$ and $\mrm Y \ra \ff $,
where $\ff$ is a hadronic system (jets), a pair of charged leptons, 
or neutrinos resulting in missing energy.
In the \general\ search mode
$\mrm X$ must be a scalar, 
$\mrm Y$ can be any scalar or vector particle of any mass, and
both particles must be short-lived so that they decay 
close to the interaction point.
The other search mode is referred to as the \ho\Zo\ search;
it requires $\mrm Y$ to be a \Zo\ boson 
and is applied to data taken at all energies.
The data used for these searches were recorded by the OPAL detector at
centre-of-mass energies (\Ecm) $88-209$~GeV, the entire energy range
achieved at LEP.

These searches are largely motivated by 
``fermiophobic'' scenarios where one of the Higgs
bosons decays primarily into a boson pair. 
In the fermiophobic interpretation, Y would be a \Zo\
and X a Higgs boson decaying into two photons. 
Indeed, the Higgs boson predicted in the Standard Model
%%% (SM)
decays into two photons via a quark- or W-boson loop~\cite{HBR},
but with a rate too low for observation of the process at LEP luminosities.
Processes $\epem \ra\hboson\Zboson \ra \gaga\ff$ 
at near-\SM\ production rate and having large di-photon branching ratios have been
predicted in a number of alternative theories~\cite{Hagiwara,Bosonic,Akeroyd,Santos,Gunion};
here \ho\ refers to the lightest neutral boson where 
extended Higgs sector models are discussed.
A particularly natural situation for fermiophobic Higgs bosons
occurs in two Higgs doublet models (2HDM)~\cite{TypeI}
of ``Type-I'', where one Higgs doublet couples only to bosons.
Because there are different fermiophobic models, it is not possible to present search results for the
entire parameter space of the various theories.
In the present paper a benchmark fermiophobic model is defined as having \SM\
production strength and a Higgs boson di-photon branching fraction calculated by
turning off the fermion couplings to the Higgs boson in the \SM.
%%%%%
%Note that, in particular fermiophobic theories, 
%interference is possible among the $\gaga$, $\WW$, and $\ZZ$ 
%final states. This interference can cause di-photon mass (\mgg) dependent structures in 
%$B({\hboson} \ra \gaga)$
%which are different from the mass behaviour in the ``benchmark'' scenario.

The OPAL Collaboration has presented searches similar to those
reported here for LEP energies up to
\Ecm\ = 189~GeV~\cite{189paper,183paper,172paper,OPAL_ggjj_1};
this paper extends those searches with the addition of data taken
at \Ecm\ = $192-209$~GeV.
Fermiophobic Higgs boson searches have also been presented
by other LEP
collaborations~\cite{ALEPHpaper,Delphipaper,L3paper}
%and these results have been combined to increase the sensitivity~\cite{LHWG}.
and by hadron collider experiments~\cite{D0paper,CDFpaper}.
To date, no evidence of a fermiophobic Higgs boson has been seen. 

%=======================================================================
\section{Data, Simulated Backgrounds and Signals}
%==============================================================================
%\subsection{Data Samples}
The data used in this analysis were recorded using
the OPAL detector~\cite{detector} at LEP.
The 1999 data consisted of $217.0\pm0.7$ \ipb\ collected at \Ecm\ = $192-202$~GeV.  
The 2000 LEP data consisted of $211.1\pm0.8$ \ipb\ collected at \Ecm\ = $200-209$~GeV,
with the majority of the data taken at 205 and 207 GeV. 
The data sets are summarized in Table~\ref{T:lumi}. 

%===============================================================================
%\subsection{Simulated Backgrounds and Signals}
%===============================================================================
%
The backgrounds from Standard Model processes were
modelled using Monte Carlo simulations at
$\sqrt{s}=192$, 196, 202, and 206~GeV for the 1999 and 2000 data.
Simulated events were processed using the full
OPAL detector Monte Carlo~\cite{GOPAL} and analysed in the 
same manner as the data.
The full-detector simulations were reweighted for the
$\sqrt{s}$ distribution of the data
using the Monte Carlo generators.

The dominant background to this search arises from the
emission of two energetic initial state radiation (ISR) photons. 
This process was simulated using
the KK2f/CEEX~\cite{CEEX} generator
with hadronisation and fragmentation by
PYTHIA 6.125~\cite{PYTHIA}.
%with the set of hadronisation parameters described 
%in Ref.~\cite{jtparams}.
The CEEX modelling of ISR employs full second-order QED corrections
to the matrix element, and applies coherent exponentiation
of the QED corrections from interference between ISR and final-state radiation. 
Four-fermion
% and $\mrm W^{\pm}e^{\mp} \nu$ 
processes were
%of the type  $\ee \ell^+ \ell^-$, where $\ell \equiv {\rm e},\mu,\tau$, were
modelled using the 
grc4f \cite{grc4f} 
and KORALW~\cite{KORALW} generators. 
Two-lepton final states were simulated using 
BHWIDE~\cite{BHWIDE}, TEEGG~\cite{TEEGG} and KORALZ~\cite{KORALZ}.
%%
%The processes $\ee\ra\ellell$ with $\ell \equiv \mu , \tau$ 
%were simulated using KORALZ~\cite{KORALZ}. 
%
The NUNUGPV~\cite{nunugpv} program was used to 
generate events of the type $\ee\ra\nunu\gamma\gamma(\gamma)$.
The process $\ee \ra \gamma\gamma$ was simulated using the
RADCOR generator~\cite{RADCOR}.
Tau lepton decays were modelled using Tauola 2.4~\cite{tauola}.

The process $\ee\ra\ho\Zo$, $\ho\ra\gamma\gamma$ 
was simulated for each \Zo\ decay channel
using the HZHA3 generator~\cite{HZHA3}. 
For the \general\ search, which is applied to the data taken in 2000 only,
the role of the \Zo\ was
replaced by scalar or vector particles having masses
from $10-200$~GeV.  
Efficiencies for signals were estimated by generating
Monte Carlo for both scalar and vector signals in mass steps of 5~GeV;
the scalar and vector efficiencies agreed within systematic errors
and therefore are not treated as separate cases. 
%In generating the
%mass grid, the recoil mass $\MY$ was varied over 
%$\mrm{E_{cm}} > \MX + \MY > M_{\mrm Z}$,
%where $\MX$ is the di-photon mass.

%=======================================================================
\section{Event Selection}
%==============================================================================
The analysis described in the following 
is identical to the one used in the paper for
OPAL data taken at 189~GeV~\cite{189paper}. 
Slightly different analysis cuts were used on the lower energy data sets, as
described in the earlier publications~\cite{183paper,172paper,OPAL_ggjj_1}.

Events were selected if there were at least two photons recoiling
from some other system decaying into one of the following
three topologies:
\begin{itemize}
\setlength{\itemsep}{-3pt}
\item[(A)] a $\qqbar$ pair (``Hadronic Channel''), or 
%%\item[(B)] a pair of oppositely charged leptons (``Leptonic Channel''), or
\item[(B)] one or two charged leptons (``Leptonic Channel''), or
\item[(C)] a $\nunu$ pair (``Missing Energy Channel'').
\end{itemize}

Photons were identified as clusters in the electromagnetic calorimeter (EC)
which were not associated with tracks
if the lateral spread of the 
clusters satisfied the criteria described in 
reference \cite{172paper}.
The efficiencies were increased by approximately $10-20$\%
by including photon conversions into \ee\ pairs
using the methods described in reference~\cite{183paper}.

The dominant background to the searches arises
from ISR 
producing mostly low-energy photons along the beam direction. 
Therefore, we required the two highest-energy photons in the event
to satisfy the following:

\begin{itemize}

 \item[(G1)] The two photon candidates were required 
       to be in the fiducial region
       $|\cos(\theta_{\gamma})| < 0.875$, 
       where the polar angle $\theta_{\gamma}$ 
       is the angle of the photon 
       with respect to the $\mrm e^-$ beam direction.

 \item[(G2)] The highest-energy photon was required to have 
       $\EgamA/\Ebeam > 0.10$ and
       the second-highest-energy photon was required to have 
       $\EgamB/\Ebeam > 0.05$.

 \item[(G3)] The sum of track momenta and 
       extra electromagnetic cluster energies
       in a 15 degree cone about the photons had to be less than 2~GeV. 

\end{itemize}

The remaining cuts depend on the particular recoil topologies.
In order to assess the background modelling, the photon cuts are not
applied until after preselection cuts for the three final state topologies.
%%%later in the analysis chain.
%The cuts for each of the three channels, as described below,
%are exactly as in Ref.~\cite{189paper}.
For all topologies, tracks and 
EC clusters 
that are not associated to tracks
are required to satisfy the criteria defined in reference \cite{CTSEL}.
%
% Charged tracks and electromagnetic clusters are subjected to several quality cuts 
%before further analysis. The tracks are required to have a transverse momentum with 
%respect to the $z$ axis greater than 0.15~GeV/$c$, a total momentum less than 90~GeV/$c$ 
%and at least 40 hits in the central jet chamber.  In addition, the distance from the
%interaction region at the point of closest approach in $r$-$\phi$ must be less than 2~cm 
%and the $z$ coordinate at this point less than 25~cm.  For clusters in the electromagnetic 
%calorimeter, the energy is required to be at least 0.10~GeV in the barrel and 0.25~GeV 
%in the endcap.  The thrust axis (see section~\ref{sec_shapes} below) is determined from 
%all tracks and electromagnetic clusters passing these cuts.  We apply a cut on the polar 
%angle $\theta_T$ of the thrust axis to ensure that the events are well contained within 
%the detector, by keeping only events with $|\costt|<0.9$.
%
%%``Unassociated'' EC clusters are defined by the requirement that
%%no track extrapolates to near the cluster.
The criteria for the definition of tracks and EC clusters in the
OPAL detector are described in reference~\cite{Zedometry}. 

%==============================================================================
\subsection{Hadronic Channel}
%==============================================================================
\label{s:qqgg}
The hadronic channel is characterised by two photons recoiling 
against a hadronic system.
Candidate events were required to satisfy the following criteria:
\begin{itemize}
\setlength{\itemsep}{-3pt}

  \item[(A1)] The standard OPAL hadronic event preselection in Ref.~\cite{hadsel};
       $\Rvis > 0.5$; $|\Sigma~p_{\mrm{z}}^{\mrm{vis}}| < 0.6 \Ebeam$;
       and at least two electromagnetic clusters with $E/\Ebeam > 0.05$.
       The quantities $\Evis$ and $\vec{p}_{\mrm{vis}}$ are the 
       scalar and vector sums of track momenta, 
       unassociated EC and unassociated hadron calorimeter
       cluster energies, and $\Rvis \equiv \frac{\mbox{\Evis}}{\Ecm}$.
       The visible momentum along the beam direction,
       obtained from the sum of all tracks and unassociated clusters,
       is denoted by 
       $|\Sigma~p_{\mrm{z}}^{\mrm{vis}}|$.

  \item[(A2)] The photon pair criteria G1$-$G3.

  \item[(A3)] Photon isolation: both photon candidates were required to satisfy
             $p_{\mrm {T,~jet}-\gamma} > 5$~GeV, where 
             $p_{\mrm {T,~jet}-\gamma}$ is the photon 
             momentum transverse to the axis of the closest jet
             out of two jets formed with the Durham~\cite{Durham} scheme 
             (excluding the photon pair).

  \item[(A4)] Photon energy balance:
             $(E_{\gamma1}-E_{\gamma2})/E_{\mrm{o}} < 0.5$, 
             where
%%%             $E_{\gamma1}$ and $E_{\gamma2}$ 
%%%             were the first and second highest energy photons
%%%             in the event, and
             $E_{\mrm{o}} \equiv (s - \MZ^{2})/(2\sqrt{s})$
             would be the energy of a single photon recoiling from the \Zo.
             This cut discriminates against ISR photon pairs.

  \item[(A5)] The recoil mass from the
              di-photon system, $\Mrec$, is required to be consistent with the $\Zo$:
              $|\Mrec - \MZ| < 20$~GeV
             (not used in the \general\ search mode).
\end{itemize}
%
%==============================================================================
\subsection{Charged Lepton Channel}
%==============================================================================
\label{s:llgg}
This channel searches for events in the $\gaga \ellell $ final state.
Events having only one well-identified lepton are accepted
to avoid efficiency loss for lepton tracks at low polar angles.
The lepton tracks are treated as jets to include tau lepton
final state topologies.
Leptonic channel candidates were required to satisfy the following
criteria:
\begin{itemize}
\setlength{\itemsep}{-3pt}

 \item[(B1)] The standard OPAL low multiplicity preselection 
             of Ref.~\cite{lowmsel}; 
            $\Rvis>0.2$; $|\Sigma~p_{\mrm{z}}^{\mrm{vis}}|<0.8 \Ebeam$;
            number of EC clusters not associated with tracks $\leq 10$;  
            number of tracks $N_{\rm T}$ satisfies  
         $1 \leq N_{\rm T} \leq 7$;
           at least two electromagnetic clusters with $E/\Ebeam > 0.05$.

  \item[(B2)] The photon pair criteria G1$-$G3.

 \item[(B3)] 
%      For events having only one track, 
%     additional requirements were made as in Ref.~\cite{189paper}.
       For events having only one track and a converted photon, the EC cluster    
       associated with the track must not also be associated with the conversion.
%     \begin{itemize} 
%       \item the track not to be associated with a converted photon;
%       \item the track to have momentum satisfying $p>0.2E_{\mrm beam}$;
%       \item direction of event missing momentum: 
%             $|\cos\theta_{\mrm miss}| > 0.90$. 
%     \end{itemize}

 \item[(B4)] For events having two or more tracks, the event
              is forced to have two jets within the Durham scheme, excluding
              the identified di-photon candidate, and
              both jets are required to have energies above 3 GeV.

  \item[(B5)] $|\Mrec - \MZ| < 20$~GeV
             (not used in the \general\ search mode).

\end{itemize} 
%

%
%==============================================================================
\subsection{Missing Energy Channel}
%==============================================================================
\label{s:nngg}
The missing energy channel is characterised by two photons 
and no other significant detector activity.
Candidates in the missing energy channel were required to 
satisfy the following criteria:
\begin{itemize}
\setlength{\itemsep}{-3pt}

\item[(C1)] The standard OPAL low multiplicity preselection of Ref.~\cite{lowmsel};
      the vetoes in Ref.~\cite{photsel} against cosmic ray and 
      beam-wall/beam-gas backgrounds; 
   number of EC clusters not associated with tracks $\leq 4$;  
   number of tracks $\leq 3$;
   $|\Sigma~p_{\mrm{z}}^{\mrm{vis}}|<0.8 \Ebeam$; and
   at least two electromagnetic clusters with $E/\Ebeam > 0.05$.

\item[(C2)] The photon pair criteria G1$-$G3.

\item[(C3)] %Consistency with the hypothesis that the di-photon system
      %is recoiling from a massive body:
      $p_T (\gaga)>0.05 \Ebeam$ 
      where $p_T (\gaga)$ is the transverse momentum
      of the di-photon system;
      the angle between the two photons in the plane 
      transverse to the beam axis: 
      $|\phi_{\gaga}-180\degree| > 2.5\degree$;
      the polar angle of the momentum of the di-photon system: 
      $|\cos\theta_{\gaga}| < 0.966$.

\item[(C4)] No track candidates other than those associated 
      with an identified photon conversion.
% as defined by the track veto criteria of reference~\cite{OPAL_HGG}. 

\item[(C5)] Veto on unassociated calorimeter energy: the energy observed 
        in the EC not associated 
        with the two photons is required to be less than 3~GeV.

\item[(C6)] $|\Mrec - \MZ| < 20$~GeV
             (not used in the \general\ search mode).
\end{itemize}

%===============================================================================
\section{Results}
\label{s:results}
%===============================================================================

For the 1999 and 2000 data, the numbers of events passing the cuts
are listed for the three recoil topologies in Table~\ref{T:gg1999-2000}.
There are no events in which more than one photon pair satisfying the
cuts was found.
The numbers of candidates passing cuts are generally in good agreement
with the expected numbers of Standard Model backgrounds;
this was also the case in earlier OPAL publications for
the lower \Ecm~\cite{189paper,183paper,172paper,OPAL_ggjj_1}. 
The one noteworthy discrepancy is for cut C1 in the missing energy
channel. In this channel there is a large background from Bhabha
electrons lost in the beampipe. The ISR photons for this background
have a steeply rising population in the forward direction. Cut C1 is
made before the cut on polar angle, and therefore a discrepancy arises
because of the steep angular distribution and the inadequate modelling
of material in the very low polar angle regions.

Combining both the 1999 and 2000 data in the three topologies,
112 candidates pass the \general\ cuts 
compared to 118.3$\pm$7.9 expected background, and
42 candidates pass the \ho\Zo  cuts 
compared to 51.9$\pm$2.9 expected background.

%===============================================================================
\subsection{Systematic Errors}
%===============================================================================
%
The uncertainty on the modelling of ISR 
is the most important component of the systematic error
because of the irreducible background arising from this process.
This uncertainty is estimated from the
comparison of data with the \SM\ background simulation
for events passing cuts A2, B2, or C2.
The shapes of the distributions for \EgamA\ and \EgamB\
%%%the higher and lower-energy photons 
are modelled well by the simulations.
The simulations also reproduce well the number of events
observed in the three channels combined.
For the 1999 and 2000 runs the simulations predict
3.7\% and 5.0\% fewer events than observed, respectively.
The statistical error on the 1999 data is 4\%,
and similarly for the 2000 data.
Modelling of photon conversions has an uncertainty of
approximately 1\%.
Uncertainties on the integrated luminosities of the
data sets are negligible compared to the other uncertainties.
Combined, these error sources result in a
total background uncertainty estimate of 10\%.
The experimental results which follow are not very sensitive 
to this number.

The dominant systematic uncertainty for the signal acceptances 
arises from the photon detection efficiency,
primarily due to the simulation 
of the photon isolation criterion G3~\cite{OPAL_ggjj_1}, 
and is estimated to be 3\%.
The uncertainty from Monte Carlo statistics
is typically better than 4\%.
A systematic error on the photon energy scale is estimated
by comparing the fitted single-photon ISR energy peak
to the expected value based
on the precisely known beam energy and \Zo\ mass.
For this study 
a sample of single-photon events was generated and compared to the
data for photon energies above 5~GeV and polar angles 
%(with respect to the beam axis) 
greater than 25 degrees.
This leads to a systematic uncertainty on the di-photon mass
of 0.35~GeV at a mass of 100~GeV.
The resolution on the di-photon mass ranges from approximately
0.5~GeV at \mgg\ = 10~GeV to 2.3~GeV at \mgg\ = 100~GeV.

%==============================================================================
\subsection{General Search Results}

\begin{figure}[!p]
        \vspace{0.8cm}
        \begin{center}
            \resizebox{\linewidth}{!}{\includegraphics{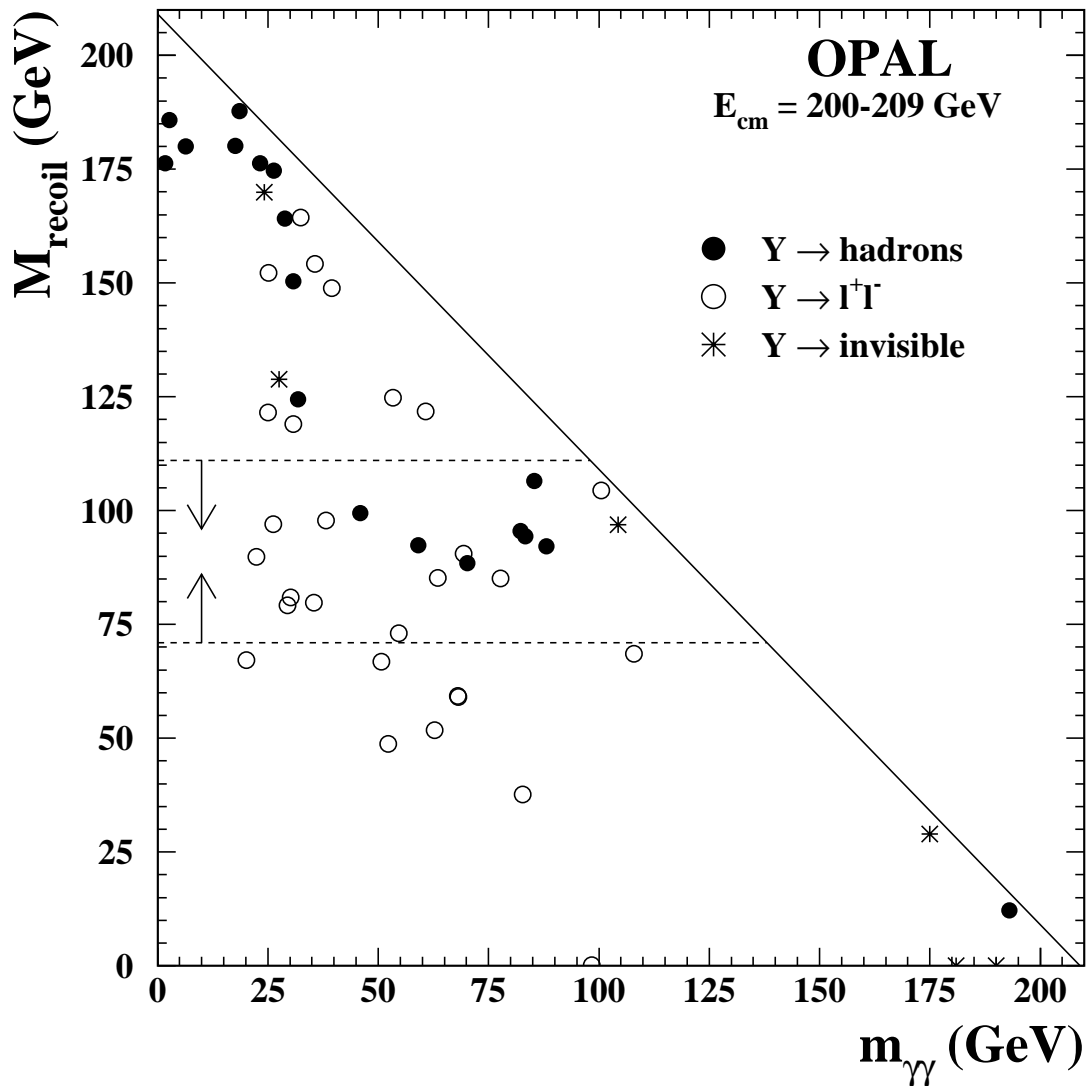} }
        \caption[COMGG2]{    
                  Distribution of mass recoiling 
                  against the di-photon system, $\Mrec$,
                  versus di-photon invariant mass, \mgg, for events passing 
                  the \general\ search cuts on the Y2000 data.
                  The different search channels are as indicated. 
                  The diagonal line denotes the kinematic limit.
                  Dashed lines and arrows indicate the events accepted
                  for the \ho\Zo\ search.
        \label{COMGG2} }
        \end{center}
    \end{figure}

Figure~\ref{COMGG2} shows the di-photon mass versus
the recoil mass for all candidate events passing the
\general\ search cuts for the year 2000 data only
(where all the data were taken at \Ecm\ near 206~GeV).
The events at recoil masses near zero are expected
from $\epem \ra \gaga$ background.
This plot also shows no unexpected structure for
the lower \Ecm\ data.
In the absence of an indication for signal,
limits are placed on the production at \Ecm$\sim$206~GeV 
of a massive state decaying into photon pairs.

For the \general\ search, the system recoiling from the di-photon system
is not assumed to be a \Zo\ and hence the branching fractions
${\mrm X} \ra \gaga$ and Y into topology A, B or C 
are not uniquely predicted.
Here X is a scalar particle
and $\mrm Y$ is a scalar or vector particle.
Furthermore, X and Y must be a short-lived particles 
so that they decay near the interaction point. 
In order to be independent of models we do not combine data
from different \Ecm\, and
therefore we restrict this part of the analysis to the 
highest energy data, in the
\Ecm\ range of $205-207$~GeV;
this represents 200.0~\ipb\ of the 2000 data.
We choose to present upper limits on
$\sigma(\epem\ra \mrm {X Y})\times B({\mrm X} \ra \gaga)\times B(\mrm Y \ra \ff)$
as a function of $\MX$.
When presenting production upper limits as functions of $\MX$,
we show the limit obtained for the value of $\MY$ that gives the smallest
efficiency in the region 
$\MZ - \MX < \MY < \mrm{E}_{\mrm{cm}} - \MX$.
The lower bounds on $\MY$ are used because
searches for di-photon resonances at LEP1~\cite{OPAL_ggjj_1,DelphiLEP1,L3LEP1}
have already set good limits on 
the lower-mass phase space.
$\MX$ and $\MY$ are also required to be  above 10~GeV and below 200~GeV in order
to allow the decay products to have sufficient energies and momenta
to give reasonable search acceptances at \Ecm\ = 206~GeV.
For a scalar/vector hypothesis for X/Y, the efficiency is found to be the
same to within 5\% as that for a scalar/scalar hypothesis;
the lower of these efficiencies is used in setting the limits.
For the lepton search channel, the efficiency for Y $\ra \tautau$ is used,
as it turns out to have the lowest of the dilepton efficiencies.

The event candidates from \Ecm\ = $205-207$~GeV in the \general\ search are
used to calculate 95\% CL upper limits on the number
of events in 1~GeV [$\MX,\MY$] mass bins.
The acceptances used at each 1~GeV mass bin are obtained
by interpolation using a 4th-order polynomial fit to the
acceptances simulated on a 5~GeV grid.
For each [$\MX,\MY$] bin, the 95\% CL upper limit  
on the number of signal events is computed using the 
frequentist method of reference~\cite{JUNK},
which takes into account the predicted \SM\ background.
This statistical procedure also incorporates the di-photon mass resolution
(typically less than 2~GeV for \mgg$<$100~GeV);
the limit procedure is valid for resonance states narrower than this resolution. 
The effect of the 10\% systematic error for background modelling
is incorporated in the statistical procedure,
as is the 4\% uncertainty on signal.
For these \general\ search limits,
an additional systematic uncertainty of 5\% is added
to the signal uncertainty
to account for interpolation error in the efficiency grid 
(especially near kinematic limits)
and for the differences in the acceptance calculations for 
the scalar versus vector nature of particle Y.

\begin{figure}[!htb]
        \vspace{0.8cm}
        \begin{center}
            \resizebox{\linewidth}{!}{\includegraphics{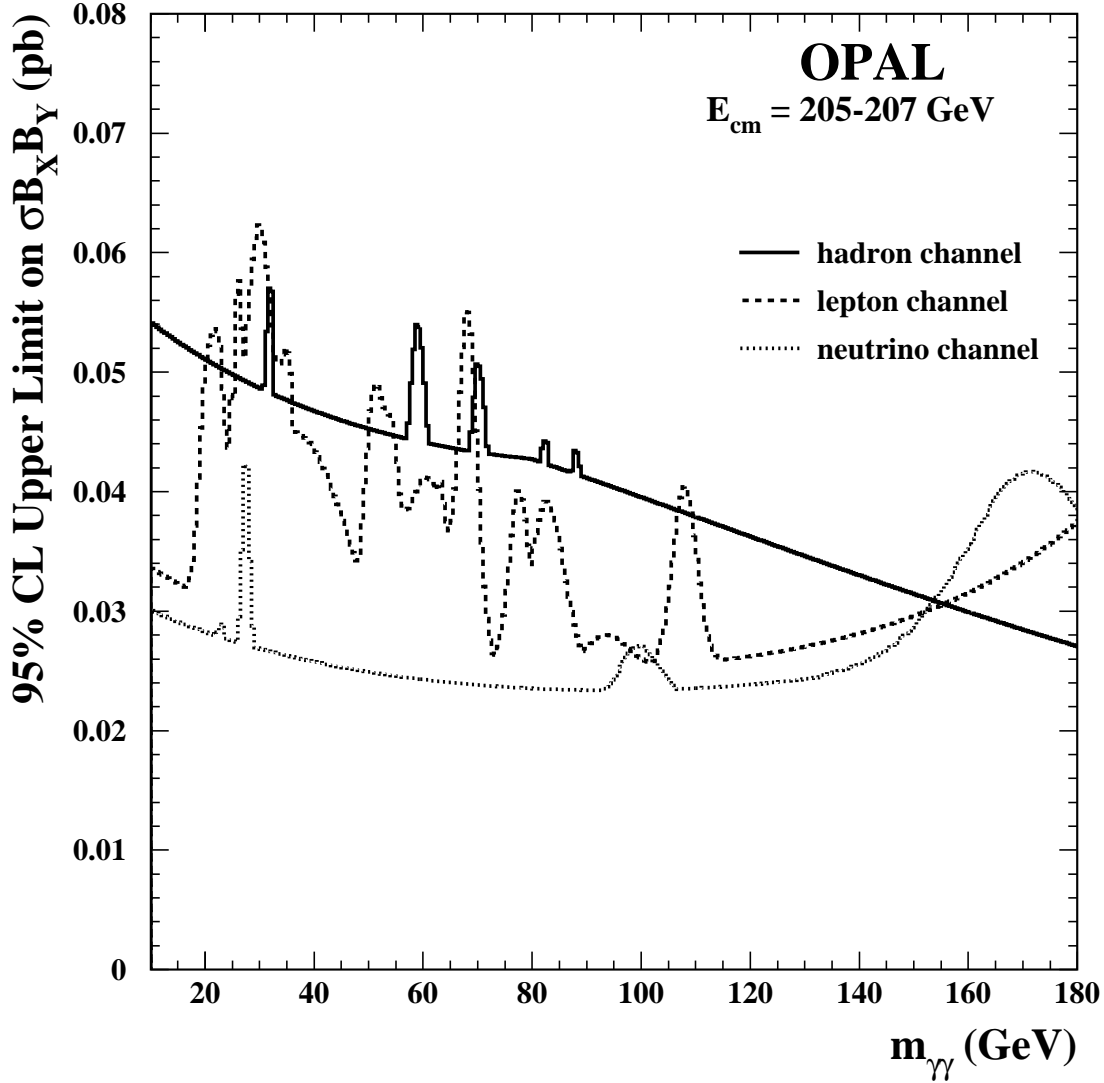} }
        \caption[limxy]{
                 For data taken at \Ecm$\sim$206~GeV: 95\% confidence level upper limit on
                 $\sigma(\epem\ra {\rm XY}) \times B(\mrm X \ra \gaga)
        \times B(\mrm Y\ra\ff)$
                 for the cases where 
                 Y decays hadronically (solid line), 
                 Y decays into a charged lepton pair (dashed line), and
                 Y decays invisibly (dotted line). 
                 The limits
                 for each $\MX$ assume the smallest efficiency as a function of
                 $\MY$ such that $10 < \MY < 200$~GeV and that
                 $\MX + \MY > \MZ$.
        \label{limxy} }
        \end{center}
    \end{figure}

Figure~\ref{limxy} shows the 95\% CL upper limits on
$\sigma(\epem\ra \mrm {X Y})\times B({\mrm X} \ra \gaga)\times 
B(\mrm Y \ra \ff)$.
These results are valid
independent of the nature of Y, provided it decays to 
%%%a fermion pair
two jets, a lepton pair, or missing energy, 
and has a width less than or equal to the experimental resolution.
Limits of $25-60$~fb are obtained 
over $10 < \MX < 180$~GeV.
The limits for the leptonic final state are stronger,
except in the case Y couples exclusively to \tautau\ 
(the final state with lowest acceptance). 

%================================================================================
\subsection{Limits on \ho\Zo\ with \ho$\ra\gaga$}

\begin{figure}[!p]
        \vspace{0.8cm}
        \begin{center}
           \resizebox{\linewidth}{!}{\includegraphics{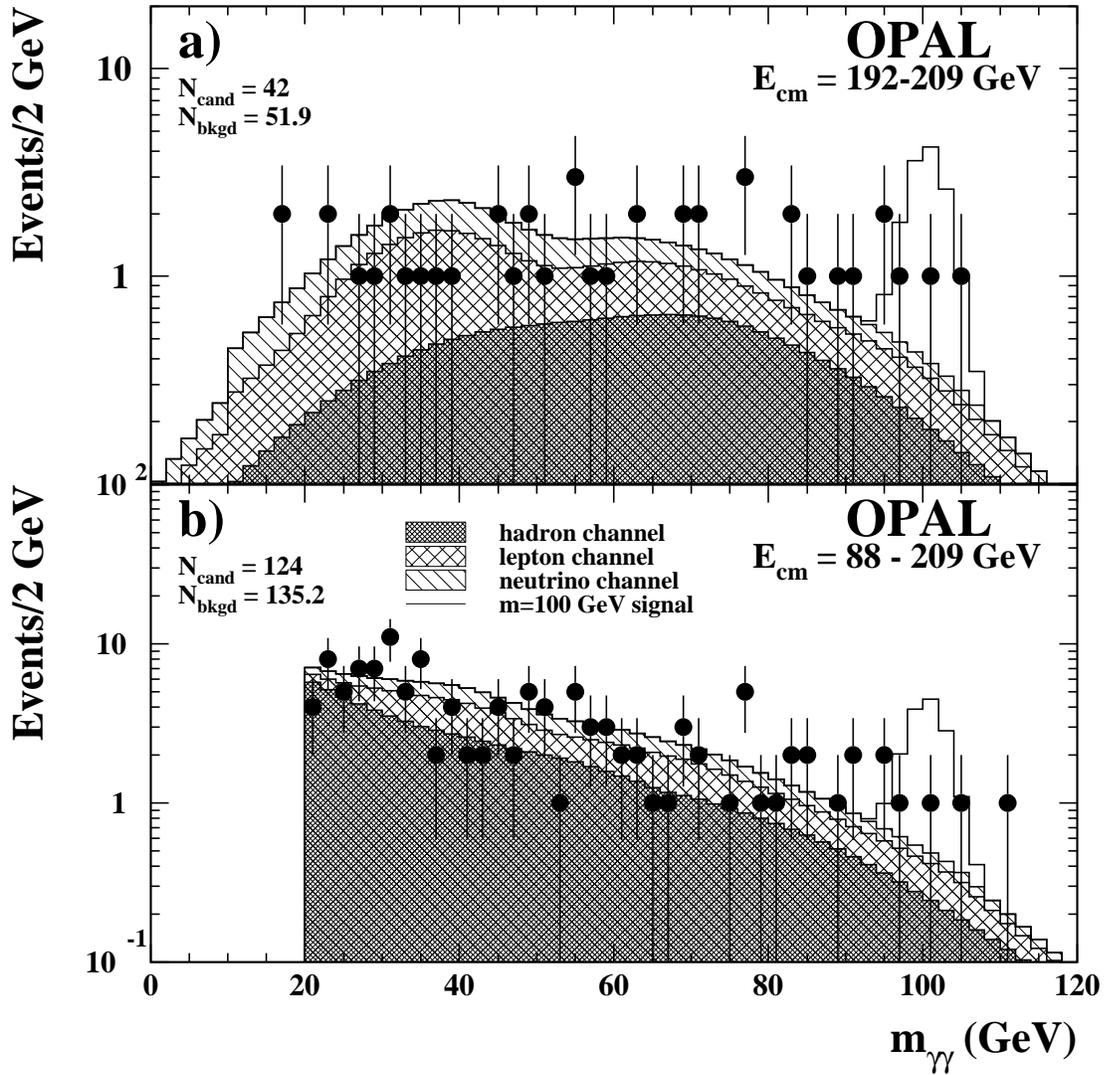} }
        \caption[COMGG]{    
                  Distribution of mass of the two highest-energy photons
                  in the \ho\Zo\ search 
                  after application of all selection criteria.
                  a) shows the data taken in 1999 and 2000 only, while 
                  b) shows the data taken at all energies.
                  Data are shown as points with error bars.
                  Background simulation is shown as a histogram 
                  showing the contributions from the hadronic, 
                  charged lepton and missing energy channels as
                  denoted.
                  The expected signal for a 100~GeV
                  fermiophobic Higgs boson produced with \SM\ strength
                  is also shown.
        \label{COMGG} }
        \end{center}
    \end{figure}

The distribution of di-photon masses for the \ho\Zo\ search candidates 
for the 
1999 data (\Ecm\ = $192-202$~GeV) and the 
2000 data (\Ecm\ = $200-209$~GeV)
is shown in Figure~\ref{COMGG}a together with the simulation of
Standard Model backgrounds. 
The observation of 42 events is
in reasonable agreement with the expected 
background of 51.9$\pm$2.9 events.
Because the \ho\Zo\ process has a production rate and branching
fractions described by theory, the data taken at all LEP energies
can be combined in this analysis.
Figure~\ref{COMGG}b shows the distribution of \mgg\ for
\Ecm\ = $88-209$~GeV. This plot is restricted to \mgg\ larger than 20~GeV because there
is no background estimate for the low-\mgg\ LEP1 data.
The figure has no indication of a resonance, and the total of
124 candidates agrees with the 
predicted background of 135.2$\pm$10.8 events.

Also indicated on Figure~\ref{COMGG} is the hypothetical signal of
a 100~GeV Higgs
boson produced at \SM\ strength and decaying into di-photons with a
branching fraction of 18\% -- the fraction predicted in the
``benchmark fermiophobic model'' calculated by simply turning off
the Higgs-fermion coupling. 
In reality, the fermiophobic Higgs photon branching ratio
depends on parameters and details of fermiophobic 2HDM models~\cite{Akeroyd,Santos},  
so this benchmark is simply a guide to 
the broad interpretation of the data.
Here we use the HDECAY~\cite{HDECAY} package to calculate the modified
photonic branching fractions. 

\begin{figure}[!htb]
        \vspace{0.8cm}
        \begin{center}
            \resizebox{\linewidth}{!}{\includegraphics{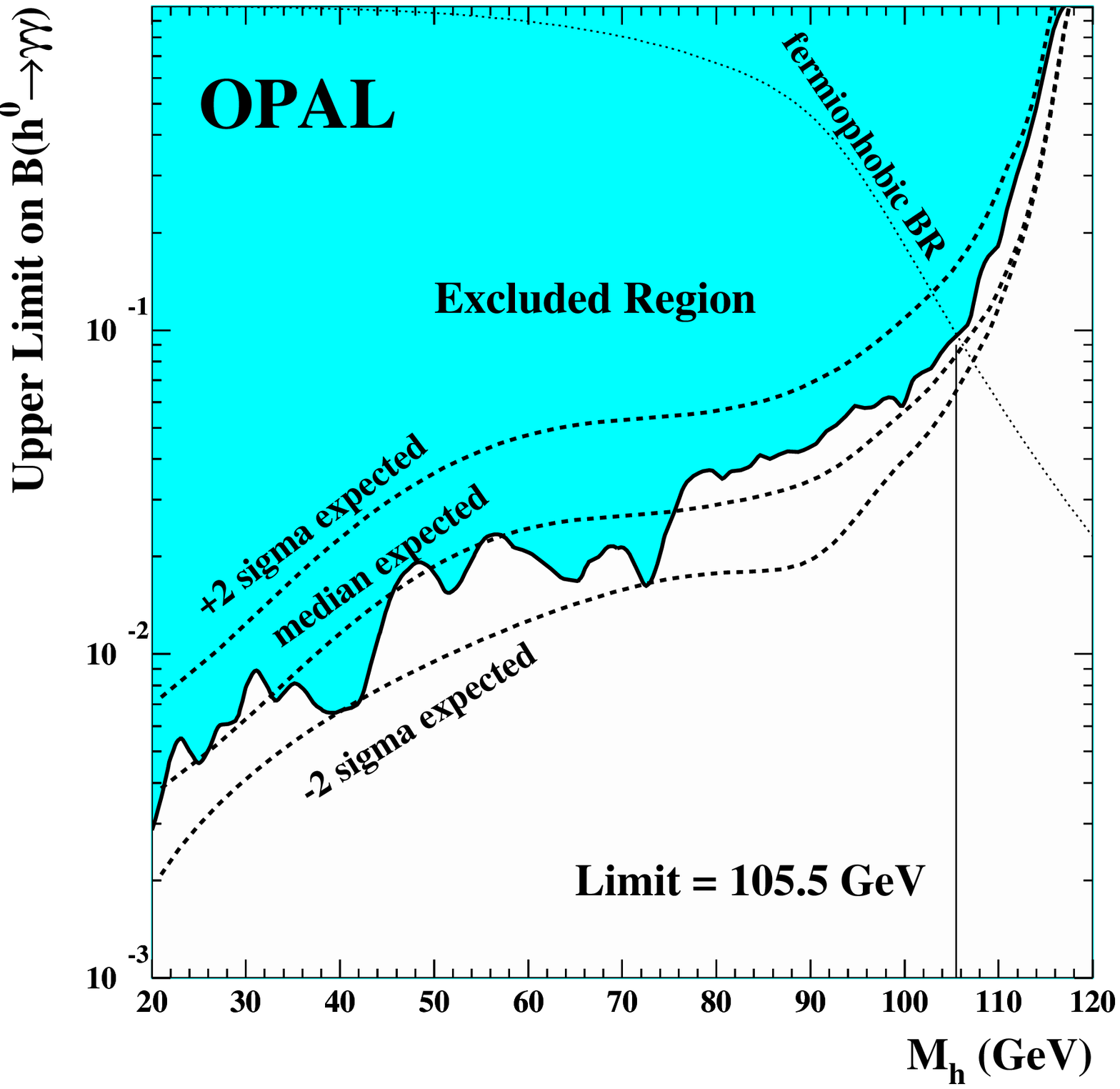} }
        \caption[bgglim]{
                 95\% confidence level upper limit on the branching fraction
                 $B$($\hboson \ra \gaga$)
                 for a Standard Model Higgs boson production rate. 
                 The shaded region, obtained with all LEP energies, is excluded. 
                 The dotted line
                 is the predicted $B$($\hboson \ra \gaga$) assuming
                 $B$($\hboson \ra \ff$) = 0. 
                 The intersection of the dotted line with the exclusion curve gives a lower limit of
                 105.5~GeV for the fermiophobic Higgs model. 
                 The median expected limit and the
        $\pm$2$\sigma$ range of expected limits are indicated by the
                 dashed lines.
        \label{bgglim} }
        \end{center}
    \end{figure}

The events passing all \ho\Zo\ cuts are used to set an upper limit
on the di-photon branching ratio for a 
particle produced in association with a \Zo\ 
and having the Standard Model Higgs boson production rate.
As described in the previous section, the frequentist method of
reference~\cite{JUNK} is used to determine the 95\% confidence level
upper limit on possible signal events at each di-photon mass.
Figure~\ref{bgglim} shows the 95\% CL upper limit for the
di-photon branching ratio 
obtained by combining the candidate events in 1999 and 2000 data described in
this paper with those from OPAL searches at 
$\sqrt{s}=88-189$~GeV~\cite{189paper,OPAL_ggjj_1,183paper},
where the Standard Model \ho\Zo\ production cross-section
is assumed at each centre-of-mass energy.
Higgs bosons produced at \SM\ rate and decaying 
exclusively to di-photons are ruled out at the 95\% confidence
level over the mass range $20-117$~GeV.
Figure~\ref{bgglim} also shows the $\ho \ra \gaga$ branching
ratio
computed using HDECAY with the fermionic couplings switched off;
the photonic branching fraction falls as the $\WW$ and $\ZZ$ channels become
kinematically favourable.
A 95\% CL lower mass limit 
for the benchmark fermiophobic Higgs bosons is set
at 105.5~GeV, where the predicted branching ratio 
crosses the upper-limit curve.  The median limit one would expect to obtain in
an ensemble of experiments in the absence of a signal is 106.4~GeV.
The benchmark fermiophobic branching ratios can also be calculated
using the HZHA3~\cite{HZHA3} generator. HZHA3 produces slightly higher
di-photon branching fractions than does HDECAY. The lower mass limit on
the benchmark fermiophobic Higgs boson calculated with HZHA3 is 106.3~GeV. 

%===============================================================================
\section{Conclusions}
%===============================================================================

A search for the production of Higgs bosons and other
new particles of width no larger than the experimental resolution and
decaying to photon pairs has been performed 
using \ee\ annihilation data with \Ecm\ = $192-209$~GeV 
combined with $88-189$~GeV data from previous OPAL searches.
Model independent upper limits are obtained for \Ecm$\sim$206~GeV on
$\sigma(\epem\ra \mrm {X Y})\times B({\mrm X} \ra \gaga)\times B(\mrm
Y \ra \ff)$,
where limits of $25-60$~fb are obtained over $10 < \MX < 180$~GeV,
for $10 < \MY < 200$~GeV and $\MX + \MY > \MZ$.
The limits are valid for Y either a scalar or vector particle, 
provided that the Y decays to a fermion pair
(interpreted as two jets, a lepton pair, or missing energy).

Using OPAL data from all LEP centre-of-mass energies,
model-specific limits are placed on
$B$($\hboson \ra \gaga$) 
up to a Higgs boson mass of 117~GeV, 
provided the Higgs particle is produced 
via $\epem \ra \hboson \Zboson$ at the Standard Model rate.
A lower mass bound of 105.5~GeV
is set at the 95\% confidence level for
benchmark fermiophobic Higgs bosons. 
Similar lower mass limits on benchmark fermiophobic Higgs bosons
have been obtained by the other LEP experiments~\cite{ALEPHpaper,Delphipaper,L3paper}.
%In combination, the LEP Higgs Working Group~\cite{LHWG} 
%has reported a preliminary
%limit of 108.2~GeV using data from the four LEP experiments. 
%%The limits obtained by the TeVatron experiments~\cite{D0paper,CDFpaper}
%%have higher mass reach, but they do not yet have the sensitivity
%%to place limits on $B$($\hboson \ra \gaga$) below approximately 40\% .
%
%
%=======================================================================
%       Acknowledgements
%=======================================================================
%\bigskip\bigskip\bigskip
%\begin{flushleft}
%{\Large\bf Acknowledgements}
%\end{flushleft}
%\par
\appendix
\par
Acknowledgements:
\par
We thank A.~G.~Akeroyd, L.~Br\"ucher, and R.~Santos for helpful
discussions.
We particularly wish to thank the SL Division for the efficient operation
of the LEP accelerator at all energies
and for their close cooperation with
our experimental group.  In addition to the support staff at our own
institutions we are pleased to acknowledge the  \\
Department of Energy, USA, \\
National Science Foundation, USA, \\
Particle Physics and Astronomy Research Council, UK, \\
Natural Sciences and Engineering Research Council, Canada, \\
Israel Science Foundation, administered by the Israel
Academy of Science and Humanities, \\
Benoziyo Center for High Energy Physics,\\
Japanese Ministry of Education, Culture, Sports, Science and
Technology (MEXT) and a grant under the MEXT International
Science Research Program,\\
Japanese Society for the Promotion of Science (JSPS),\\
German Israeli Bi-national Science Foundation (GIF), \\
Bundesministerium f\"ur Bildung und Forschung, Germany, \\
National Research Council of Canada, \\
Hungarian Foundation for Scientific Research, OTKA T-029328, 
and T-038240,\\
Fund for Scientific Research, Flanders, F.W.O.-Vlaanderen, Belgium.\\

%=======================================================================
%       References
%=======================================================================

%%%%%%%%%%%%%%%%%%%%%%%%%%%%%%%%%%%%%%%%%%%%%%%%%%%%%%
% result tables
%%%%%%%%%%%%%%%%%%%%%%%%%%%%%%%%%%%%%%%%%%%%%%%%%%%%%%
\newpage
%
%===============================================================================
% lumi table
%===============================================================================
\begin{table}[htbp]
\begin{center}

%%{\footnotesize

\begin{tabular}{|l|c|c|}\hline

Run (year) &Integrated Luminosity (\ipb) &\Ecm\ (GeV) \\ \hline

1990-95  &173.00  &$88-94$ \\ \hline
1995     &5.41    &$130-140$ \\ \hline
1996     &10.32   &$172.3$ \\ \hline
1996     &10.04   &$161.3$ \\ \hline
1997     &57.73   &$182.6$ \\ \hline
1998     &182.61  &188.6 \\ \hline
%\multicolumn{2}{|c|}{Year 1999} \\ \hline
1999     &28.90 &191.6 \\ \hline
1999     &74.79 &195.6 \\ \hline
1999     &77.21 &199.6 \\ \hline
1999     &36.08 &201.6 \\ \hline
%\multicolumn{2}{|c|}{Year 2000}\\ \hline
2000     &0.82 &$200-202$ \\ \hline
2000     &2.62 &$202-204$ \\ \hline
2000     &76.81 &$204-206$ \\ \hline
2000     &123.26 &$206-208$ \\ \hline
2000     &7.54 &$208-210$ \\ \hline
\hline
\end{tabular}
%%}
\end{center}
  \caption[Integrated Luminosity]
   {Summary of all data sets used in the searches.
    Results using data from $1990-1998$ were reported in earlier 
    publications~\cite{189paper,183paper,172paper,OPAL_ggjj_1}.}
  \label{T:lumi}
\end{table}
%
%============================================================================

%===============================================================================
% qq,ll,nunu table
%===============================================================================
\def\mulcA{\multicolumn{1}{|c|}}
\def\mulcC{\multicolumn{3}{|c||}{}}
\begin{table}[htbp]
\begin{center}

{\footnotesize

\begin{tabular}{|l||r||r|r|r|r|r|r||r|}\hline
 Cut  & Data & \mulcA{$\Sigma$Bkgd} & \mulcA{$(\gamma/{\rm Z})^{\ast}$} & \mulcA{4f} 
 &\mulcC
 &\mulcA{$\epsilon_{100}$ (\%)} \\ 
\hline
\multicolumn{9}{|c|}{1999 Hadronic Channel }\\ \hline
 (A1)    & 10645 & 10695.4      &  7535.0  &  3160.2 &\mulcC &70  \\ \hline
 (A2)    & 62    & 56.1         &  53.6    &  2.6    &\mulcC &62  \\ \hline
 (A3)    & 48    & 44.8         &  42.5    &  2.3    &\mulcC &61  \\ \hline
 (A4)    & 29    & $22.0\pm1.8$ &  19.8    &  2.2    &\mulcC &61  \\ \hline \hline
 (A5)    & 15    & $10.9\pm0.9$ &  10.9    &  0.0    &\mulcC &60  \\ \hline
\multicolumn{9}{|c|}{2000 Hadronic Channel }\\ \hline
 (A1)    & 9371  & 9152.4       &  6096.0  &  3056.2 &\mulcC &71  \\ \hline 
 (A2)    & 52    & 46.5         &  43.5    &  3.0    &\mulcC &60  \\ \hline
 (A3)    & 39    & 38.3         &  36.0    &  2.3    &\mulcC &60  \\ \hline
 (A4)    & 18    & $17.8\pm1.6$ &  15.7    &  2.1    &\mulcC &58  \\ \hline \hline
 (A5)    & 7     & $7.9\pm0.7$  &  7.8     &  0.1    &\mulcC &57  \\ \hline
\multicolumn{9}{c}{} \\ \hline

 Cut   & Data   &\mulcA{$\Sigma$Bkgd} &\mulcA{$\ee$} & \mulcA{$\tautau$}  
                    & \mulcA{$\mm$} & \mulcA{$\gaga$}  
                               & \mulcA{$\ee\ff$}  
 &\mulcA{$\epsilon_{100}$ (\%)} \\ 
\hline
\multicolumn{9}{|c|}{1999 Leptonic Channel }\\ \hline
 (B1)  & 41947 & 39060.6   & 37245.9 & 715.6 & 50.5 & 314.6 & 734.0 &82  \\ \hline
 (B2)  & 167   & 188.6     & 72.4  & 11.3  & 8.0  & 95.5  & 1.5     &69  \\ \hline
 (B3)  & 155   & 178.8     & 67.3  & 10.4  & 7.4  & 92.4  & 1.2     &63  \\ \hline
 (B4)  & 23  & $31.4\pm5.8$& 18.8  & 5.0   & 6.9  & 0.4   & 0.3         &50  \\ \hline \hline
 (B5)  & 5   &$9.2\pm1.7$  & 4.6   & 1.7   & 2.9  & 0.0   & 0.0     &48  \\ \hline
\multicolumn{9}{|c|}{2000 Leptonic Channel }\\ \hline
 (B1)  & 37432 & 33928.9    & 32329.1 & 607.8 & 43.7 & 273.9 & 674.4 &82  \\ \hline
 (B2)  & 138   & 146.6      & 57.6    & 10.2  & 6.8  & 70.9  & 1.0   &71  \\ \hline
 (B3)  & 123   & 141.7      & 55.4    & 9.2   & 6.4  & 69.7  & 1.0  &67  \\ \hline
 (B4)  & 28    & $23.7\pm4.8$ & 13.0    & 4.4   & 5.9  & 0.2   & 0.3  &61  \\ \hline \hline 
 (B5)  & 11    &$9.8\pm2.0$ & 5.1     & 1.9   & 2.6  & 0.0   & 0.0   &52  \\ \hline
\multicolumn{9}{c}{} \\ \hline

 Cut  & Data & \mulcA{$\Sigma$Bkgd} & \mulcA{$\nunu\gaga$} & \mulcA{$\gaga$} 
           & \mulcA{$\ee$} & \mulcA{$\ell^+\ell^-$} & \mulcA{$\ee\ff$}  
 &\mulcA{$\epsilon_{100}$ (\%)} \\ 
\hline
\multicolumn{9}{|c|}{1999 Missing Energy Channel }\\ \hline
 (C1)   & 224989 & 129713.4 & 50.5 & 3337.7 & 124769.9 & 157.4 & 1397.9 &88  \\ \hline
 (C2)   & 377  & 336.3  & 13.0 & 276.4  & 45.1   & 1.1   & 0.7          &75  \\ \hline
 (C3)   & 71   & 73.1   & 12.2 & 31.5   & 27.8   & 1.0   & 0.5          &69  \\ \hline
 (C4)   & 33   & 42.4   & 12.1 & 29.2   & 0.9    & 0.0   & 0.2          &69  \\ \hline
 (C5)   & 8    & $12.9\pm0.5$   & 11.4 & 0.3  & 0.9    & 0.0   & 0.2            &67  \\ \hline \hline
 (C6)   & 3    &$7.6\pm0.3$ & 7.5 & 0.0   & 0.0    & 0.0   & 0.1        &66  \\ \hline
\multicolumn{9}{|c|}{2000 Missing Energy Channel }\\ \hline
 (C1)   & 202649 & 113268.8   & 43.9 & 2737.6 & 109070.9& 136.8 &1279.7 &87  \\ \hline 
 (C2)   & 345    & 315.1      & 11.5 & 267.5  & 34.6    & 0.9   & 0.5    &75  \\ \hline
 (C3)   & 66     & 62.1       & 11.0 & 30.1   & 19.7    & 0.8   & 0.3    &70  \\ \hline
 (C4)   & 34     & 39.0       & 11.0 & 27.8   & 0.0     & 0.0   & 0.2    &69  \\ \hline
 (C5)   & 6      & $10.5\pm0.5$  & 10.2 & 0.1    & 0.0     & 0.0   & 0.1    &68  \\ \hline \hline
 (C6)   & 1      &$6.5\pm0.3$ & 6.5  & 0.0    & 0.0     & 0.0   & 0.0    &65  \\ \hline

\end{tabular}

}
\end{center}
  \caption[Data and MC after cuts]
  { The $\ho\ra\gamma\gamma$ searches:
   events from the 1999 and 2000 data 
   remaining after the indicated cumulative cuts.
   Hadronic channel cuts (A),
   leptonic channel cuts (B),
   and missing energy channel cuts (C) are described in Sections 
\ref{s:qqgg}, \ref{s:llgg}, and \ref{s:nngg}, respectively. 
   The entries for A5, B5, C6 are for the $\Mrec$ cut for the \ho\Zo\ search;
   they are not applied in the \general\ search.
   The uncertainty shown indicates statistical error only;
   the systematic uncertainty is approximately 10\%.
    In (A), the components from
   $(\gamma/{\rm Z})^{\ast}$ and four-fermion (``4f") final states
   are shown. 
  In (B), the components from
   Bhabha scattering ($\ee$),
   $\tau$-pair, $\mu$-pair, $\gaga$ and $\ee\ff$ final states 
   are shown. 
   In (C), the components from
   $\nunu\gaga$, $\gaga$, $\ee$-pair, lepton pair ($\ell\equiv\mu,\tau$) 
   production and $\ee\ff$ final states are shown. 
   The efficiencies for detection of a 100~GeV Higgs boson are shown
   in the last column.
}
  \label{T:gg1999-2000}
\end{table}
%
%============================================================================

%%%%%%%%%%%%%%%%%%%%%%%%%%%%%%%%%%%%%%%%%%%%%%%%%%%%%%%%%%%%%%%%%%%%%%%%
% figures 
%%%%%%%%%%%%%%%%%%%%%%%%%%%%%%%%%%%%%%%%%%%%%%%%%%%%%%%%%%%%%%%%%%%%%%%%
%*********************************************************************
\newpage

%*********************************************************************
\newpage

%*********************************************************************
\newpage

%*********************************************************************
\newpage

%%%%%%%%%%%%%%%%%%%%%%%%%%%%%%%%%%%%%%%%%%%%%%%%%%%%%%%%%%%%%%%%%%%%%%%%
\end{document}